\documentstyle[epsfig,twocolumn]{iso98}

\newcommand{\thepapername}{}

\begin{document}

\setlength{\parindent}{0pt}
\setlength{\parskip}{ 10pt plus 1pt minus 1pt}
\setlength{\hoffset}{-1.5truecm}
\setlength{\textwidth}{ 17.1truecm }
\setlength{\columnsep}{1truecm }
\setlength{\columnseprule}{0pt}
\setlength{\headheight}{12pt}
\setlength{\headsep}{20pt}
\pagestyle{veniceheadings}

\title{ISOPHOT OBSERVATIONS OF COMET HALE-BOPP: INITIAL DATA REDUCTION
\thanks{ISO is an ESA
project with instruments funded by ESA Member States (especially the PI
countries: France, Germany, the Netherlands and the United Kingdom) and
with the participation of ISAS and NASA.}}

\author{{\bf E. Gr\"un$^{1}$, S. B. Peschke$^{1}$, M. Stickel$^{2}$, T. M\"uller$^{2,3}$, H.
Kr\"uger$^{1}$, H. B\"ohnhardt$^{4}$, T. Y. Brooke$^{2}$,}\\  
{\bf H. Campins$^{6}$, J. Crovisier$^{7}$, M. S. Hanner$^{5}$, I. Heinrichsen$^{3}$, H. U. Keller$^{8}$,
R. Knacke$^{9}$,} \\
{\bf P. Lamy$^{10}$, Ch. Leinert$^{2}$, D. Lemke$^{2}$,
C. M.Lisse$^{11}$, M. M\"uller$^{1,12}$, D. J. Osip$^{6}$, M. Solc$^{13}$,} \\
\vspace{5mm}
{\bf M. Sykes$^{14}$, V. Vanysek$^{13,+}$ and J. Zarnecki$^{15}$}\\
$^{1}$Max-Planck-Institut f\"ur Kernphysik, Heidelberg, Germany \\
$^{2}$Max-Planck-Institut f\"ur Astronomie, Heidelberg, Germany \\
$^{3}$ISO Data Center, ESA, Villafranca, Spain \\
$^{4}$European Southern Observatory, Santiago, Chile \\
$^{5}$Jet Propulsion Laboratory, Pasadena, USA \\
$^{6}$University of Florida, Gainesville, USA \\
$^{7}$Observatoire de Paris, Meudon, France \\
$^{8}$Max-Planck-Institut f\"ur Aeronomie, Katlenburg-Lindau, Germany \\
$^{9}$Penn State University, Erie, USA \\
$^{10}$Laboratoire d'astrophysique spatial, Marseille, France \\
$^{11}$University of Maryland, College Park, USA \\
$^{12}$ESA-ESOC, Darmstadt, Germany \\
$^{13}$Charles University, Prague, Czech Republic \\
$^{14}$University of Arizona, Tucson, USA \\
$^{15}$University of Kent, Canterbury, UK \\
$^{+}$deceased
}
\maketitle

\begin{abstract}

Comet Hale-Bopp was observed five times with ISOPHOT, the photometer on
board the Infrared Space Observatory (ISO). Each time broadband
photometry was performed using 4 different detectors, 5 apertures and,
10 filters covering the range between 3.6 to 175\,$\mu$m. Calibration
measurements using the internal Fine Calibration Source were done
together with each individual measurement. Background observations were
performed with identical instrument settings at the same positions on
the sky several days after the comet observations. The observation
strategy and the initial data reduction steps are described in some
detail and the resulting in-band power values of the Hale-Bopp
observations and their uncertainties are derived. Initial reduction of
these measurements was performed in 3 steps: (1) processing of raw data
by removing instrumental and energetic particle effects, (2) averaging of
the individual signals, and (3) determination of the detector
responsivities and their uncertainties. The detector signal is determined by
two different methods and the best value is chosen. At the present level
of processing uncertainties range from 10\% to a factor of 3 
for the low power levels at short wavelengths. 
The in-band power levels at different wavelengths
varied over 3 orders of magnitude.

\end{abstract}

\section{OBSERVATIONS}

Comet Hale-Bopp was a unique target-of-oppor\-tunity comet during the 
lifetime of the ISO satellite. Photometric observations of comet Hale-Bopp
were performed with the ISOPHOT instrument (Lemke et al., 1996). The aim
of this paper is to describe the observations and the initial data
reduction scheme in enough detail in order to understand the present
uncertainties of the measurement results. Astronomical interpretation of
the results in terms of the spectral energy distribution requires
additional color and aperture corrections and will be reported in a
separate paper (Gr\"un et al., in preparation). ISO viewing constraints 
together with
the comet's orbital geometry limited the visibility of Hale-Bopp for ISO.
Observations were only allowed in the solar elongation range of
$60^{\circ}$ to $120^{\circ}$. In addition, the Earth had to be at least
$24^{\circ}$ away from the satellite axis. Further pointing constraints
were imposed by the Moon and Jupiter. Because of these constraints,
comet Hale-Bopp was only observable for ISO during two periods in spring
and autumn 1996 and again in winter 1997. Hale-Bopp's perihelion passage
occurred in April 1997.

Within the visibility period of Hale-Bopp the lowest possible background
had to be found in front of which the observations were to be performed.
The search for the lowest background along the arc of Hale-Bopp's orbit
was done in two subsequent steps. First, a raw selection based on the
IRAS all-sky survey (ISSA) plates was performed. 'Dark regions' at 12,
25, 60 and 100$\mu$m along the cometary track on the ISSA plates were
selected. Second, the selected regions were examined with IRSKY
(IPAC) to assure that no known point sources were within 5 arcmin of the
cometary track, and no prominent (cirrus) structures were visible on the
100$\mu$m ISSA plates. Also the background estimates for each filter of
the ISO observations were derived with IRSKY. 

Background subtraction is crucial for photometric measurements but
standard simultaneous offset measurements were dismissed because of the
extended coma of Hale-Bopp. Instead we introduced so-called 'shadow
measurements'. A shadow measurement is the repetition of the Hale-Bopp
measurement, called master observation, at the same position on the sky
that was tracked in the master observation, at a time when the comet and
more importantly the coma had moved away from that sky position. Shadow
observations were typically performed about one week after the master
observations. For this kind of background measurements we assumed that
the seasonal variations of the zodiacal emission are small. Because of
scheduling problems the 3rd Hale-Bopp observation was not executed as
planned but was delayed by about one month and, therefore, these master
and  shadow observations are a few degrees apart. 

Four different detectors of ISOPHOT (Lemke et al., 1996) were used for
the observations (Table~\ref{hale-bopp-table}): detector P1, a Si:Ga-detector for
observations up to 16\,$\mu$m, detector P2, a Si:B-detector for the
25\,$\mu$m observations, the C100 camera, a 3x3-pixel-array of nine
Ge:Ga-detectors for 60 and 100\,$\mu$m observations, and the C200
camera, a 2x2 array of four stressed Ge:Ga-detectors for the 175\,$\mu$m
observations. All observations were done in single pointing, single
filter absolute photometry mode (AOT P03 and P22), i.e. with a
subsequent measurement of the Fine Calibration Source (FCS). 
Multi-aperture measurements (AOT P04) of comet Halo-Bopp are 
reported in a companion paper (Peschke et al., 1998).  

\section{DATA PROCESSING}

Processing ISOPHOT data from raw data to meaningful astronomical values
is a multi-step process with several decisions to be made that affect
the end result. This process is described in some detail in order to get
an estimate of the uncertainties of the result, and, to allow other
researchers to compare their results with ours. For the initial data
reduction - that is described in this paper - the ISOPHOT Interactive
Analysis tool (PIA, Gabriel et al. 1997) is used as a baseline. 
In principal, it takes care of detector and instrumental effects and
finally   results in in-band power values for each individual comet and
background   measurement (per detector pixel and per bandpass).  The
second step  that comprises color corrections, 
corrections for flux
deficits due  to inaccurate pointing and correct aperture scaling   
for the distributed coma brightness will
be discussed later (Gr\"un et al., in preparation).  All these
corrections are needed to derive a spectral energy distribution in a
standardized  aperture over a wide wavelength range.

The five ISOPHOT Hale-Bopp observation sequences reported here comprise
558 individual measurements, counting all object, shadow, and
calibration (FCS) measurements at up to 10 wavelengths, some of which
were multi-pixel measurements with up to 9 pixels. Each of these
measurements had to undergo all initial steps of data processing
described below before they were combined into 44 Hale-Bopp and
background signal values (Table~\ref{hale-bopp-table}). 

During the exposure to an infrared source the ISOPHOT detector output
voltage ramps-up with time at a rate that is a function of the infrared
flux received by the detector. This voltage is recorded and read-out in
preset time steps of 1/128 s to 1/4 s. The output voltage is digitized
with 12 bits resolution over a dynamic range of about 2.2 V. Normally,
before the detector output voltage reaches saturation, after a fixed
number of read-outs, the voltage is reset and a new voltage ramp is
started. Each individual measurement consisted of 4 to 683 voltage
ramps. That way, from 512 to 4096 individual voltage values per
measurement were obtained. The individual read-out programs were defined
in the observation planning phase dependent on the expected flux levels.
Table~\ref{hale-bopp-table} shows parameters of the different measurements. 

 The objective of the initial data processing is to derive the flux 
dependent slope of the voltage ramps [V/s] and to determine the
responsivity  of the detector.  The whole data was reduced homogeneously
with PIA version 7.1.  At raw data (ERD) level, non-linearities of the
detector response  and  of the electronic amplification are corrected by
using the   algorithms provided by PIA.   During the whole data
reduction process, the default criteria for discarding  data points have
been applied. In case of only 4 readouts per ramp, just the
two middle data points are left for the slope determination of the ramp.

 Next, the effects of high energetic particle hits are removed which
cause the signal  to display discrete steps of variable amplitudes on
top of the linear rise.  These glitches are removed in two steps, partly
on the raw data level, partly  after the determination of the slopes.
The slopes of the integration ramps have been determined 1) by first
order fits to   the ramps (method~1) and ,
2) by determining the in pairs-differences  of subsequent
non-destructive read-outs (method~2).  By applying the drift recognition algorithm
with its default settings,   an average signal for the measurement was
deduced.

 To get confidence in the values derived with these two different
methods within   PIA, a third independent method was introduced to have
an independent handle on   the effect of glitch removal as 
well as on detector drifts.   For this, only the non-linearity 
correction was applied to the raw data and further processing occurred
outside PIA.  All in pairs-differences were computed, regardless
of the type of readout. At this stage, the overall trend in the data,
the effect of detector drifts  and glitches was clearly visible in the
data. Glitches could easily be identified  by their distance to the bulk
data, based on very good statistics.  Next, a myriad algorithm was
applied to the data which computed  and  analyzed the distribution of
slope values. A trim fraction   was introduced to remove the
tails of the distribution that contain mainly glitches.  Then the
remaining data is analyzed for its

\begin{table*}[p]
  \caption{\em Log of Hale-Bopp and corresponding shadow observations. For
each individual ISOPHOT measurement, detector, wavelength of filter,
aperture size and integration time are given. For some observations 
the complete set of filters could not be used. For the specific 
measurements (Hale-Bopp and shadow) the number of voltage ramps, number 
of read-outs per ramp and the resulting responsivity and in-band power 
values are given
}
  \begin{center}
\label{hale-bopp-table}
    \leavevmode
    \footnotesize
    \begin{tabular}[h]{cccc|cccc|cccc}
\hline                          
Detector & Filter & Aper-    & Inte-       & \# of & \# of     & Respon- & In-band  & \# of & \# of 
& Respon-& In-band \\
         &        &  ture    & gration     & ramps & read-outs & sivity  & power & ramps & read-outs 
& sivity & power \\
         &        &  size    &  times      &       & per ramp  &         &        &       & per ramp 
         &     &        \\
         &[$\mu$m]& ["]      &   [s]       &       &             &   [A/W] & [W]    &      &    
&  [A/W]   &  [W]   \\
\hline 
  \multicolumn{4}{c|}{observations}                             
& \multicolumn{4}{|c|}{Hale-Bopp, 25-MAR-1996}                             
& \multicolumn{4}{|c}{Shadow,  30-MAR-1996}\\                           
\hline 
P1 & 3.6  & 23  & 256 & 128 & 8   & 1.18  & $4.1\cdot 10^{-16}$  & 128 & 8  & 1.18  &    $2.7\cdot 10^{-16}$            \\
P1 & 7.3  & 23  & 32  & 128 & 4   & 1.18  & $3.8\cdot 10^{-15}$  & 128 & 4  & 1.18  &    $1.5\cdot 10^{-15}$            \\
P1 & 10   & 52  & 32  & 128 & 4   & 1.18  & $2.8\cdot 10^{-14}$  & 128 & 4  & 1.75  &    $5.0\cdot 10^{-15}$            \\
P1 & 12.8 & 52  & 32  & 128 & 4   & 1.18  & $2.6\cdot 10^{-14}$  & 128 & 4  & 3.60  &    $3.4\cdot 10^{-15}$            \\
P1 & 16   & 52  & 32  & 128 & 4   & 1.62  & $1.5\cdot 10^{-14}$  & 128 & 4  & 1.18  &    $6.0\cdot 10^{-15}$            \\
C1 & 60   & 135 & 32  & 4   & 256 & 70.13 & $6.2\cdot 10^{-14}$  & 32  & 32 & 71.25 &    $1.3\cdot 10^{-14}$            \\
C1 & 100  & 135 & 32  & 4   & 256 & 70.51 & $2.9\cdot 10^{-14}$  & 32  & 32 & 54.38 &    $1.2\cdot 10^{-14}$            \\
C2 & 175  & 180 & 32  & 8   & 128 & 21.59 & $2.5\cdot 10^{-14}$  & 16  & 64 & 20.59 &    $1.8\cdot 10^{-14}$            \\
\hline                          
  \multicolumn{4}{c|}{observations}                             
& \multicolumn{4}{|c|}{Hale Bopp,  27-APR-1996}                         
& \multicolumn{4}{|c}{Shadow,  5-MAY-1996}\\                             
\hline                                  
P1 & 3.6  & 23  & 256 & 128 & 8   & 1.18  & $1.5\cdot 10^{-16}$  & 128 & 8  & 1.17  &    $2.7\cdot 10^{-17}$            \\
P1 & 7.3  & 23  & 32  & 128 & 4   & 1.18  & $3.0\cdot 10^{-15}$  & 128 & 4  & 1.17  &    $7.1\cdot 10^{-16}$            \\
P1 & 10   & 52  & 32  & 128 & 4   & 1.18  & $3.3\cdot 10^{-14}$  & 128 & 4  & 1.17  &    $3.5\cdot 10^{-15}$            \\
P1 & 12.8 & 52  & 32  & 128 & 4   & 1.18  & $3.1\cdot 10^{-14}$  & 128 & 4  & 1.17  &    $5.2\cdot 10^{-15}$            \\
P1 & 16   & 52  & 32  & 128 & 4   & 1.53  & $1.8\cdot 10^{-14}$  & 128 & 4  & 1.17  &    $3.2\cdot 10^{-15}$            \\
P2 & 25   & 99  & 32  & 64  & 32  & 0.78  & $3.0\cdot 10^{-13}$  & 128 & 16 & 0.64  &    $4.5\cdot 10^{-14}$            \\
C1 & 60   & 135 & 32  & 4   & 256 & 32.51 & $7.1\cdot 10^{-14}$  & 32  & 32 & 33.70 &    $9.1\cdot 10^{-15}$            \\
C1 & 100  & 135 & 32  & 4   & 256 & 42.49 & $2.8\cdot 10^{-14}$  & 32  & 32 & 33.84 &    $1.1\cdot 10^{-14}$            \\
C2 & 175  & 180 & 32  & 8   & 128 & 22.08 & $2.4\cdot 10^{-14}$  & 32  & 32 & 20.55 &    $1.5\cdot 10^{-14}$            \\
\hline                         
  \multicolumn{4}{c|}{observations}                             
& \multicolumn{4}{|c|}{Hale Bopp,  27-SEP-1996}                         
& \multicolumn{4}{|c}{Shadow, 6-SEP-1996}\\                             
\hline                                  
P1 & 3.6  & 23  & 256 & 128 & 8   & 1.18  & $2.3\cdot 10^{-15}$  & 128 & 8  & 1.16  &    $1.5\cdot 10^{-16}$            \\
P1 & 7.3  & 23  & 32  & 128 & 8   & 1.18  & $1.0\cdot 10^{-13}$  & 128 & 4  & 1.16  &    $8.8\cdot 10^{-16}$            \\
P1 & 10   & 52  & 32  & 128 & 16  & 1.18  & $4.3\cdot 10^{-13}$  & 128 & 4  & 1.16  &    $4.1\cdot 10^{-15}$            \\
P1 & 12.8 & 52  & 32  & 64  & 32  & 1.18  & $2.4\cdot 10^{-13}$  & 128 & 4  & 1.16  &    $5.3\cdot 10^{-15}$            \\
P1 & 16   & 52  & 32  & 64  & 32  & 1.18  & $2.3\cdot 10^{-13}$  & 128 & 4  & 1.16  &    $3.4\cdot 10^{-15}$            \\
C1 & 60   & 135 & 32  & 4   & 342 & 38.20 & $1.8\cdot 10^{-13}$  & 64  & 16 & 36.23 &    $6.5\cdot 10^{-15}$            \\
C1 & 100  & 135 & 32  & 4   & 342 & 39.54 & $8.3\cdot 10^{-14}$  & 16  & 64 & 31.93 &    $1.6\cdot 10^{-14}$            \\
C2 & 175  & 180 & 32  & 4   & 256 & 22.19 & $6.1\cdot 10^{-14}$  & 16  & 64 & 26.52 &    $5.6\cdot 10^{-14}$            \\
\hline                         
  \multicolumn{4}{c|}{observations}                             
& \multicolumn{4}{|c|}{Hale Bopp,  7-OCT-1996}                         
& \multicolumn{4}{|c}{Shadow,  10-OCT-1996}\\                            
\hline                                  
P1 & 3.6  & 23  & 256 & 128 & 8   & 1.18  & $1.8\cdot 10^{-15}$  & 128 & 8  & 1.16  &    $1.9\cdot 10^{-17}$            \\
P1 & 7.3  & 23  & 32  & 64  & 32  & 1.15  & $8.7\cdot 10^{-14}$  & 128 & 4  & 1.16  &    $1.1\cdot 10^{-15}$            \\
P1 & 10   & 52  & 32  & 32  & 64  & 1.18  & $2.7\cdot 10^{-13}$  & 128 & 4  & 1.16  &    $5.0\cdot 10^{-15}$            \\
P1 & 12.8 & 52  & 32  & 32  & 64  & 1.02  & $2.0\cdot 10^{-13}$  & 128 & 4  & 1.16  &    $6.5\cdot 10^{-15}$            \\
P1 & 16   & 52  & 32  & 32  & 64  & 1.38  & $1.7\cdot 10^{-13}$  & 128 & 4  & 1.16  &    $3.9\cdot 10^{-15}$            \\
P2 & 25   & 99  & 32  & 8   & 256 & 0.60  & $2.4\cdot 10^{-12}$  & 128 & 16 & 0.72  &    $5.2\cdot 10^{-14}$            \\
C1 & 60   & 135 & 32  & 4   & 342 & 39.29 & $1.5\cdot 10^{-13}$  & 64  & 16 & 36.34 &    $6.2\cdot 10^{-15}$            \\
C1 & 100  & 135 & 32  & 4   & 342 & 39.11 & $7.3\cdot 10^{-14}$  & 32  & 32 & 33.07 &    $1.1\cdot 10^{-14}$            \\
C2 & 175  & 180 & 32  & 4   & 256 & 22.81 & $6.3\cdot 10^{-14}$  & 16  & 64 & 26.46 &    $3.8\cdot 10^{-14}$            \\
\hline                          
  \multicolumn{4}{c|}{observations}                             
& \multicolumn{4}{|c|}{Hale Bopp,  30-DEC-1997}                             
& \multicolumn{4}{|c}{ Shadow, 4-JAN-1998}\\                            
\hline                                  
P1 & 3.6  & 23  & 256 & 128 & 8   & 1.18  & $2.9\cdot 10^{-16}$  & 128 & 8  & 1.18  &    $2.6\cdot 10^{-18}$            \\
P1 & 7.3  & 23  & 256 & 128 & 32  & 1.18  & $7.3\cdot 10^{-15}$  & 128 & 8  & 1.18  &    $3.5\cdot 10^{-16}$            \\
P1 & 10   & 52  & 64  & 128 & 32  & 1.18  & $4.4\cdot 10^{-14}$  & 128 & 4  & 1.18  &    $2.0\cdot 10^{-15}$            \\
P1 & 11.3 & 52  & 64  & 64  & 64  & 0.89  & $8.1\cdot 10^{-14}$  & 128 & 4  & 1.18  &    $4.9\cdot 10^{-15}$            \\
P1 & 12.8 & 52  & 64  & 128 & 32  & 1.14  & $4.0\cdot 10^{-14}$  & 128 & 4  & 1.18  &    $2.6\cdot 10^{-15}$            \\
P1 & 16   & 52  & 64  & 128 & 32  & 1.36  & $3.6\cdot 10^{-14}$  & 128 & 4  & 1.18  &    $1.4\cdot 10^{-15}$            \\
P2 & 25   & 99  & 64  & 16  & 256 & 0.76  & $5.2\cdot 10^{-13}$  & 128 & 8  & 0.97  &    $1.3\cdot 10^{-14}$            \\
C1 & 60   & 135 & 64  & 4   & 683 & 25.97 & $8.3\cdot 10^{-14}$  & 64  & 32 & 38.49 &    $1.5\cdot 10^{-15}$            \\
C1 & 100  & 135 & 64  & 4   & 512 & 32.60 & $4.3\cdot 10^{-14}$  & 64  & 32 & 36.85 &    $2.2\cdot 10^{-15}$            \\
C2 & 175  & 180 & 64  & 16  & 128 & 23.25 & $2.4\cdot 10^{-14}$  & 64  & 32 & 24.68 &    $7.1\cdot 10^{-15}$            \\
      \hline \\
      \end{tabular}
  \end{center}
\end{table*}

momentum and a stable mean method  is
applied. The derived value reflects the average signal of the
measurement. This  value together with the appropriate error is written back
into a structure for   further processing within PIA.

Method 3 is
superior to the previous methods when the measurement comprises only a small number
of ramps. We calculate the slopes by all methods in parallel but use
method 3 for the final result whenever possible. However, in case the
voltage increase per read-out is less than 3 mV method 3 deteriorates
because of digitization noise and results from method 1 or 2 are used.
Independent of the method by
which the average signal slope was derived, the standard reset 
interval correction  is
applied, dark current is subtracted and the effect of vignetting is
corrected  in case of the C detectors.

\begin{figure}[bth]
   \begin{center}
   \leavevmode
   \centerline{\epsfig{file=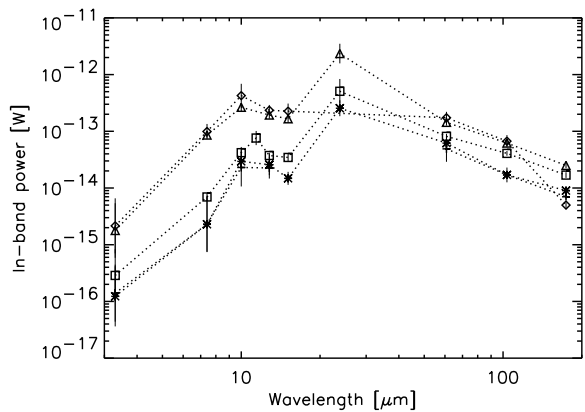,
               width=8.0cm}}
   \end{center}
 \caption{\em In-band power values of the five Hale-Bopp observations.
Different symbols characterise different observations: 
$+$:		25-MAR-1996,
$\ast$:	27-APR-1996,
$\diamond$:	27-SEP-1996,
$\triangle$:	7-OCT-1996,
$\Box$:		30-DEC-1997.
Two measurements at 25$\mu$m wavelengths were not completed 
because of spacecraft problems.
  }
 \label{fig-hale-bopp}
 \end{figure}

All the above processing steps are applied both to the object and the
subsequent FCS measurements. In an FCS
measurement the detector is irradiated by the internal calibration lamp
(FCS) and the actual detector responsivity can be determined. However,
this method works only reliably when the infrared power onto the
detector from the FCS is within about a factor of 3 of the power from
the object. The FCS power was set beforehand on the basis of estimates
of the object brightness. In case of larger deviations - this was the
case in about 2/3 of the measurements - the default responsivity should
give more reliable results. The actual and default responsivities
sometimes showed large deviations: up to a factor of 3 and more,
especially in the case of the very low power values at short 
wavelengths. In
several of these cases the FCS heating power and the corresponding power
of infrared radiation were out of the calibrated range. In any case, we
use the deviation of the different responsivity values (default,
 and actual responsivities determined by all three methods) 
as a measure of the uncertainty of
the responsivity and the derived in-band power value.
Figure~\ref{fig-hale-bopp} shows the derived in-band power values of the 
Hale-Bopp measurements. 
For a single observation the in-band powers at different
wavelengths vary over about 3 orders of magnitudes.
Note especially the close similarity of the in-band power
values of the 27-SEP-1996 and 7-OCT-1996 observations, 
except for the value at 175$\mu$m. Due to scheduling problems
the positions of the master and the shadow observations were offset 
by a few degrees which only showed a strong effect at the longest wavelength.

\section*{Acknowledgments}
Effective support by the ISO Project and ISOPHOT Teams is acknowledged. This
research was supported by a DLR grant.

\end{document}